\documentclass[a4paper,10pt,twoside]{cpc-hepnp}
\usepackage{multicol}
\usepackage{graphicx}
\usepackage{booktabs}
\usepackage{amssymb,bm,mathrsfs,bbm,amscd}
\usepackage[tbtags]{amsmath}
\usepackage{lastpage}
\begin{document}
\fancyhead[co]{\footnotesize L. Girlanda et al: Electromagnetic processes in a $\chi$EFT framework}
\footnotetext[0]{Received 14 December 2009}
\title{Electromagnetic processes in a $\chi$EFT framework \thanks{DOE DE-AC05-06OR23177 }}
\author{%
           L. Girlanda$^{1,2;1)}$\email{luca.girlanda@pi.infn.it}%
\quad S. Pastore$^{3;2)}$\email{pastore@jlab.org}%
\quad R. Schiavilla$^{3,4;3)}$\email{schiavil@jlab.org}%
\quad M. Viviani$^{2;4)}$\email{michele.viviani@pi.infn.it}%
}
\maketitle
\address{%
1~(Department of Physics, University of Pisa, 56127 Pisa, Italy)\\
2~(INFN-Pisa, 56127 Pisa, Italy)\\
3~(Department of Physics, Old Dominion University, Norfolk, VA 23529, USA)\\
4~(Jefferson Lab, Newport News, VA 23606)\\
}
\begin{abstract}
Recently, we have derived a two--nucleon potential and
consistent nuclear electromagnetic currents in
chiral effective field theory with pions and nucleons
as explicit degrees of freedom. The calculation
of the currents has been carried out to include N$^3$LO
corrections, consisting of two--pion exchange
and contact contributions. The latter involve
unknown low-energy constants (LECs), some of which
have been fixed by fitting the $np$ S- and P-wave
phase shifts up to 100 MeV lab energies. The remaining
LECs entering the current operator are determined so
as to reproduce the experimental deuteron and trinucleon
magnetic moments, as well as the $np$ cross section.
This electromagnetic current operator is utilized to study
the $nd$ and $n^3$He radiative captures
at thermal neutron energies.
Here we discuss our results stressing on the important role played
by the LECs in reproducing the experimental data.
\end{abstract}

\begin{multicols}{2}
Quantum chromodynamics (QCD) is the underlying theory of the strong
interaction. On this basis, interactions among the relevant degrees
of freedom of nuclear physics, such as pions, nucleons, and delta-isobars,
are completely determined by the quark and gluon dynamic.
At low energies though, the strong coupling constant becomes too large
to allow for application of  perturbative techniques to solve QCD. Consequently,
we are still far from a quantitative understanding of the low-energy
physics by {\it ab initio} calculations from QCD. Chiral effective field theory
($\chi$EFT) exploits the symmetries exhibited by QCD in the low-energy
regime, in particular chiral symmetry, to constrain the form of the interactions
of the pions among themselves and with the other degrees of
freedom\cite{Weinberg95}. The pion couples by powers of its momentum $Q$
and the Lagrangians describing these interactions can be expanded
in powers of $Q/\Lambda_\chi$, where $\Lambda_\chi\sim1$ GeV
represents the chiral-symmetry breaking scale and characterizes
the convergence of the expansion.  The effectiveness of the theory is then
confined to kinematic regions where the constraint $Q\ll \Lambda_\chi$ is realized.
The unknown coefficients of the chiral expansion, {\it i.e.} the low energy
constant (LECs), need to be fixed by comparison with the experimental data.
$\chi$EFT provides an expansion of the Lagrangians in powers of a small
momentum as opposed to an expansion in the strong coupling constant,
restoring {\it de facto} the applicability of perturbative techniques also in the
low-energy regime. Due to the chiral expansion it is possible, in principle,
to evaluate an observable to any degree of desired accuracy and
to know {\it a priori} the hierarchy of interactions contributing to
the low energy process under study.

Since the pioneering work of Weinberg\cite{Weinberg90},
this calculational scheme has been widely utilized in nuclear physics and nuclear
$\chi$EFT has developed into an intense field of research. Nuclear
two-- and three--body interactions\cite{Epelbaum08}, as well as interactions of
electroweak probes with nuclei\cite{Park93,Park96} have been studied within
the $\chi$EFT approach. 

Recently, we have derived the nuclear electromagnetic (EM)
currents in $\chi$EFT\cite{Pastore08,Pastore09}, retaining, as degrees of freedom,
pions and nucleons.  The calculation has been carried out in time-ordered
perturbation theory\cite{Pastore08} with non-relativistic Hamiltonians
derived from the chiral Lagrangians of Refs.~\citep{Weinberg90,vanKolck94,Epelbaum98}.
The strong and electromagnetic interaction Hamiltonians required to evaluate the EM current operator up to
N$^3$LO accuracy---that is $e\, Q$ in the chiral expansion, $Q$ denoting the low momentum
scale, and $e$ being the electric charge---are listed in Ref.~\citep{Pastore08,Pastore09}.

In Fig.~\ref{fig:fig1} we show the contributions to the current operator up to
N$^2$LO ($e\,Q^0$). The LO ($e\,Q^{-2}$) term is given by a one-body contribution, consisting of
the standard convection and spin-magnetization nucleon currents, while
pion-exchange currents occur at NLO ($e\,Q^{-1}$). The N$^2$LO term is due to
($Q/M$)$^2$ relativistic corrections---where $M$ denotes the nucleon mass---to the
LO one-body current. 

In Fig.~\ref{fig:fig2} we list the N$^3$LO contributions, 
which can be separated into three classes: i)
one-loop two-pion exchange terms, represented
by diagrams (a)-(i); ii) tree-level term involving
the nuclear-electromagnetic Hamiltonian of order $e\,Q^2$ at the vertex illustrated
by a full circle in diagram (j); and iii) contact currents of minimal and non-minimal nature,
illustrated by diagram (k). 

The last two contributions involve unknown LECs.
In particular, the tree-level current of the type shown in panel (j),
depends on three LECs, two of them multiply isovector structures and the
remaining one multiplies an isoscalar structure. Incidentally, the isovector part of
this tree-level current has the same  structure as the current involving the excitation
of a delta-isobar\cite{Pastore08}. This resonance saturation argument is exploited to infer the
ratio between the two LECs multiplying the isovector terms in the current of diagram (j) (see below).
Contact currents of non-minimal character, panel (k) in Fig.~\ref{fig:fig1},
depend on two additional unknown LECs, multiplying respectively an isoscalar and
an isovector structure, while those obtained via minimal substitution are expressed in terms
of LECs entering the contact two-nucleon chiral potential of order
$Q^2$ (or N$^2$LO)~\cite{Pastore09}.  The two-nucleon potential
has been derived in Ref.~\citep{Pastore09} up to N$^2$LO and these LECs have been fixed by
fitting the $np$ S- and P-wave phase shifts up to 100 MeV laboratory energies\cite{Pastore09}.
Thus total number of unknown LECs to be determined is reduced to four.

\begin{center}
\includegraphics[width=6cm]{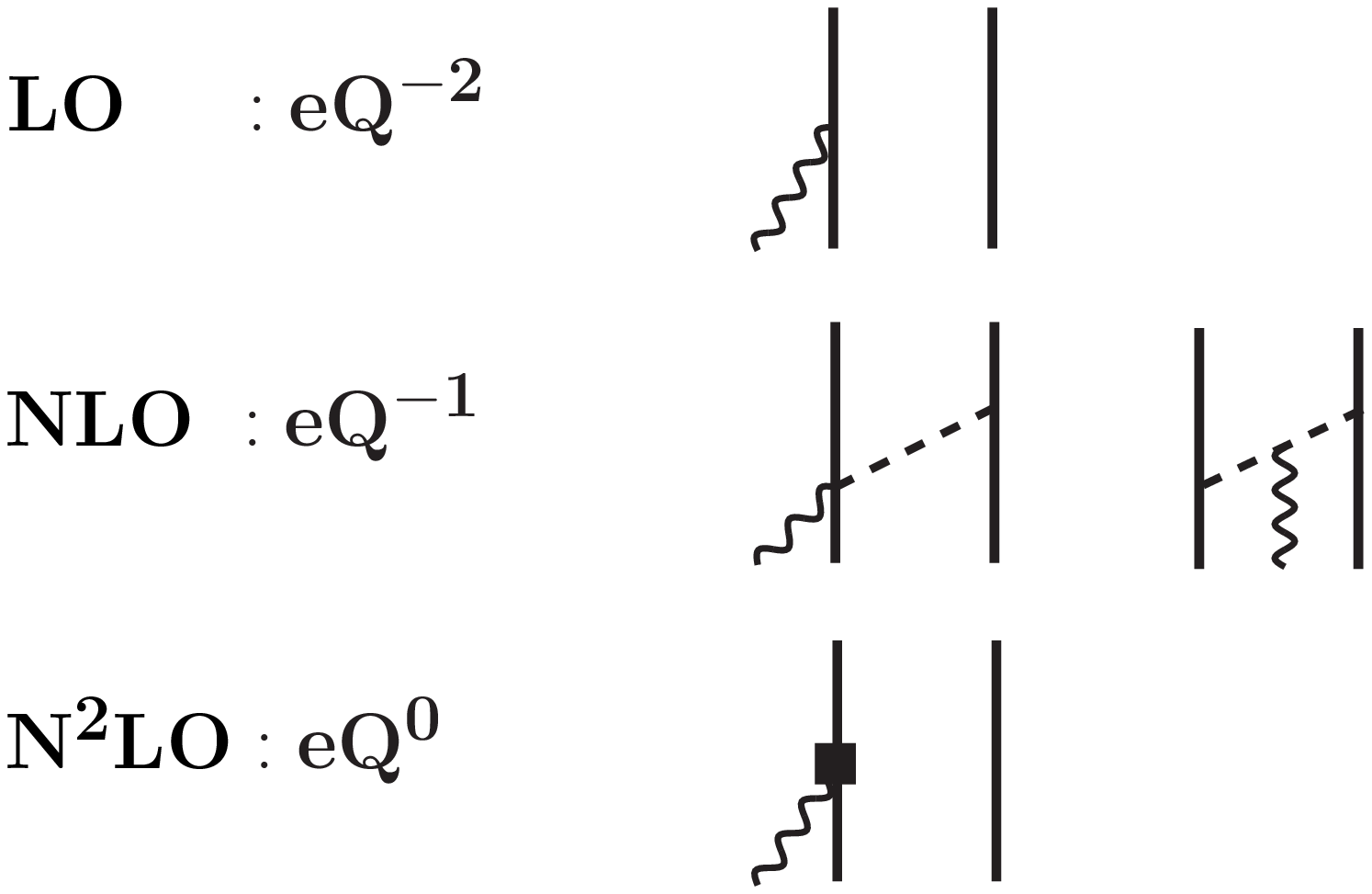}
\figcaption{\label{fig:fig1}   {Diagrams illustrating one- and two-body currents up to N$^2$LO ($e\, Q^{0}$).
Nucleons, pions, and photons are denoted by solid, dashed, and wavy lines,
respectively. The square represents the relativistic correction to the
LO one-body current. Only one among the possible time orderings is shown for the NLO diagrams.} }
\end{center}

\end{multicols}
\ruleup
\begin{center}
\includegraphics[width=12cm]{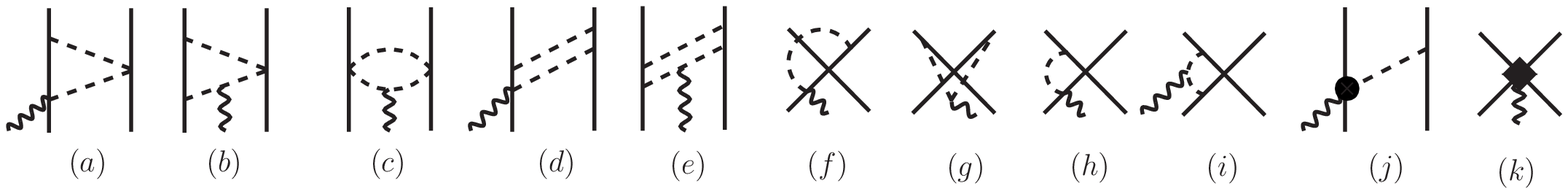}
\figcaption{\label{fig:fig2} Diagrams illustrating two-body currents entering at N$^3$LO
($e\, Q$), notation as in Fig.~\protect\ref{fig:fig1}.
Only one among the possible time orderings is shown for diagrams (a)-(j).}
\end{center}
\ruledown
\begin{multicols}{2}

An important aspect of the derivation
of the EM currents (and two-nucleon potential) is to
retain both irreducible diagrams and recoil-corrected
reducible ones\cite{Pastore08}.
The latter arise from expanding the energy denominators (in reducible
diagrams) in powers
of nucleon kinetic energy differences to pion energies (these ratios are
of oder $Q$).  Partial cancellations occur between the irreducible and
recoil-corrected reducible contributions
both at N$^2$LO and N$^3$LO\cite{Pastore08}.
We also note that this approach leads to N$^3$LO EM currents
that satisfy the continuity equation
with the corresponding N$^2$LO two-body potential\cite{Pastore08}.
The expressions for the two-pion-exchange N$^3$LO currents
in panels (a)-(i)
of Fig.~\ref{fig:fig2} are in agreement with those obtained by
K\"olling {\it et al.} in Ref.~\citep{Koelling09} by the method of the unitary
transformations.  However, they are different from those derived
by Park {\it et al.} in Ref.~\citep{Park96} in covariant
perturbation theory, since these authors include irreducible contributions
only.

We now present a study of the $nd$ and $n^3$He radiative capture
at thermal neutron energies within the hybrid approach, where the
EM $\chi$EFT current operator described above is used to evaluate
transition matrix elements between nuclear wave functions obtained
with realistic Hamiltonian with two-- and three--body potentials.
In order to study the model dependence of the calculated observables,
we use two different combinations of two-- and three--body potential, namely
the Argonne $v_{18}$\cite{Wiringa95} with the
Urbana-IX\cite{Pudliner97} three--nucleon potential (AV18/UIX), and the
N$^3$LO\cite{Entem03} and N$^2$LO\cite{Gazit09} chiral two--
and three--nucleon potentials (N3LO/N2LO).
We study the sensitivity of the observables to variations of
the cutoff $\Lambda$, introduced to regularize the
EM current operator via the momentum cutoff
$C_\Lambda(k)\!\!=$\,exp$(-k^4/\Lambda^4)$. In our study,
$\Lambda$ varies from $500$ to $700$ MeV which corresponds to
``removing'' short-range physics at distance scales less
$1/(3\, m_\pi)$.

\vspace{0.55cm}
\begin{center}
\includegraphics[width=6cm]{Fig3.eps}
\figcaption{\label{fig:fig3}   {Cumulative LO, NLO, N$^2$LO, and N$^3$LO(S-L) contributions
for the deuteron and trinucleon isoscalar and isovector magnetic moments, and $np$ radiative capture. } }
\end{center}

Out of the four unknown LECs entering the EM current operator,
two multiply isoscalar structures and two multiply isovector
operator structures.  We fix these LECs by reproducing the experimental values
of two isoscalar observables, {\it i.e.}~the deuteron [$\mu_d$] and the
isoscalar [$\mu^S$($^3$He/$^3$H)]
combination of the trinucleon magnetic moments, and
two isovector observables, {\it i. e.} the isovector [$\mu^V$($^3$He/$^3$H)]
combination of the trinucleon magnetic moments and
the $np$ cross section [$\sigma_{np}^\gamma$] at thermal neutron energies.
The results are shown in Fig.~\ref{fig:fig3} where the cumulative contributions
at LO, NLO, N$^2$LO, and N$^3$LO(S-L) are represented.
The cumulative contribution N$^3$LO(S-L) is given by
the terms up to N$^2$LO plus the N$^3$LO contributions
associated with pion loops (represented in panels (a)-(i)
of Fig.~\ref{fig:fig2}), which depend on the (known)
nucleon axial coupling constant, pion decay amplitude, and pion mass, as well 
as with contact currents, which depend on the
LECs obtained from the fits to the $np$ phase shifts.

The LECs entering the complete current, denoted in what follows as N$^3$LO(LECs),
are fixed, for each value of the cutoff $\Lambda$, so as to reproduce the experimental
values which in Fig.~\ref{fig:fig3} are represented by the black band, including
experimental errors.  The sensitivity of the results to the two Hamiltonian models utilized
(AV18/UIX and the N3LO/N2LO) is represented
by the thickness of the color bands.
We note that the sign of the N$^2$LO and N$^3$LO(S-L) contributions
is opposite to that of the LO and NLO contributions. This increases
the discrepancy between theory and experiment.

\vspace{0.9cm}
\begin{center}
\includegraphics[width=6cm]{Fig4.eps}
\figcaption{\label{fig:fig4}  { Cumulative LO, NLO, N$^2$LO, N$^3$LO(S-L), and N$^3$LO(LECs) 
contributions to the $nd$ ($\sigma_{nd}^\gamma$) 
and $n^3$He ($\sigma_{n^3{\rm He}}^\gamma$) cross sections (right and left top panel respectively),
and circular polarization factor $R_c$.}}
\end{center}

Having fixed all the LECs, we are left with a completely determined
EM current operator which can now be used to make predictions for the
$n$($d$,$\gamma$)$^3$H and $n$($^3$He,$\gamma$)$^4$He
reactions' cross sections---denoted as $\sigma_{nd}^\gamma$ and
$\sigma_{n^3{\rm He}}^\gamma$ respectively---and the circular polarization
factor $R_c$ associated with the capture of polarized neutrons on deuterons.
In this calculation we have used the AV18/UIX (N3LO/N2LO) combination of two- and
three-nucleon potentials for the $A=$3 ($A=$4) processes; calculations with the N3LO/N2LO
(AV18/UIX) potential models are in progress.
The predictions are represented in Fig.~\ref{fig:fig4} along with the experimental
data, shown in black, which are from Ref.~\citep{Jurney82} for $nd$
and Ref.~\citep{Wolfs89} for $n^3$He.
The complete N$^3$LO(LECs) current is shown in Fig.~\ref{fig:fig4}
by the orange lines.
The calculated $nd$ cross section is in excellent agreement
with the measured value and is weakly dependent on the cutoff.
The cross section for the $n\, ^3$He reaction undergoes a 5\%
variation when the cutoff changes from $500$ to $700$ MeV, but is still
within the experimental error band. These reactions are known
to be dominated by many-body components of the current operator,
which provide most of the calculated cross section\cite{Carlson98}.
This trend is confirmed here: the LO
contribution to the cross sections is highly suppressed, and provides only
about $46\%$ ($18\%$) of the total calculated $nd$ ($n^3{\rm He}$) value.
What is more interesting though, is the large contribution associated with 
the N$^3$LO(LECs) currents in both these reactions.  These currents
are crucial for bringing theory into agreement with experiment. 

We are presently in the process of extending these hybrid
studies to different realistic Hamiltonian
models, with the goal of quantifying the sensitivity of the cross sections
to the wave functions employed in the calculations.  Obviously,
our ultimate objective is to perform a fully
consistent $\chi$EFT calculation, using the N$^2$LO potential
derived in Ref.~\citep{Pastore09}, along with the EM currents we
presented here.  In Ref.~\citep{Pastore09} we
show the deuteron wave functions obtained with the N$^2$LO
chiral potential and compare them with those
corresponding to the AV18.  The two sets of wave functions
display a different behavior at short range, in particular
the N$^2$LO D-wave component is significantly smaller
than the AV18.  From this perspective, it will be interesting to establish whether
these chiral potential and currents lead to a satisfactory
description of the $nd$ and $n\,^3$He captures.
\\

\acknowledgments{We thank J.L.\ Goity and R.B.\ Wiringa for useful
discussions. The work of R.S.\ is supported by the U.S.~Department of Energy,
Office of Nuclear Physics, under contracts DE-AC05-06OR23177.
Some of  the calculations were made possible by grants of computing
time from the National Energy Research Supercomputer Center.}

\end{multicols}

\vspace{-2mm}
\centerline{\rule{80mm}{0.1pt}}
\vspace{2mm}

\begin{multicols}{2}

\end{multicols}

\vspace{5mm}

\clearpage

\end{document}